\documentclass[preprint,prd,aps,nofootinbib]{revtex4}
\usepackage{amsmath}
\usepackage{latexsym}
\usepackage{psfig}
\usepackage{graphicx}
\usepackage{subfigure}
\usepackage[pdfauthor={Alexander Dolgov, Diego N. Pelliccia},pdftitle=
{Scalar field instability in de Sitter space-time},pdfsubject=
{\langle\varphi^{2}\rangle}, pdfkeywords={scalar field, quantum
field theory in curved space, instabilities, de Sitter,
fluctuations}]{hyperref}
\newcommand{\be}{\begin{equation}}
\newcommand{\ee}{\end{equation}}
\newcommand{\ba}{\begin{eqnarray}}
\newcommand{\ea}{\end{eqnarray}}
\newcommand{\bc}{\begin{center}}
\newcommand{\ec}{\end{center}}
\def\cstok#1{\leavevmode\thinspace\hbox{\vrule\vtop{\vbox{\hrule\kern1pt
\hbox{\vphantom{\tt/}\thinspace{\tt#1}\thinspace}}
\kern1pt\hrule}\vrule}\thinspace}
\begin{document}
\title{Scalar field instability in de Sitter space-time}
\author{Alexander Dolgov$^{1,2,3}$
\thanks{Electronic address: dolgov@fe.infn.it},
Diego N. Pelliccia$^{1,2}$
\thanks{Electronic address: diego.pelliccia@fe.infn.it}}
\affiliation{ ${ }^{1}$Istituto Nazionale di
Fisica Nucleare, Sezione di Ferrara,\\
Via Paradiso 12, 44100 Ferrara, Italy\\
${ }^{2}$Universit\`{a} di Ferrara,
Dipartimento di Fisica,\\
Via Paradiso 12, 44100 Ferrara, Italy\\
${ }^{3}$ITEP,
Bol. Cheremushkinskaya 25, Moscow 113259 Russia\\
}

\vspace{2cm}
\begin{abstract}

Starting from the equation of motion of the quantum operator of a
real scalar field $\varphi$ in de Sitter space-time, a simple
differential equation is derived which describes the evolution of
quantum fluctuations $\langle\varphi^{2}\rangle$ of this field.
Full de Sitter invariance is assumed and no \emph{ad hoc} infrared
cutoff is introduced. This equation is solved explicitly and in
massive case our result agrees with the standard one. In massless
case the large time behavior of our solution differs by sign from
the expression found in earlier papers. A possible cause of
discrepancy may be a spontaneous breaking of de Sitter
invariance.

\end{abstract}

\maketitle

\section{Introduction}

Since de Sitter found his exact solution to Einstein's equations
in 1917, this model of space-time has been intensively studied
because of its maximal symmetry. Moreover, according to
inflationary scenarios the universe was in (quasi) de Sitter state
at some part of its history. Most probably inflation was induced
by a scalar field (inflaton) and this fact amplifies the interest
to the study of scalar fields in de Sitter space. Recent
observations of Supernovae of type Ia \cite{Riess}, implying that
present universe is in an accelerated stage, refreshed the
interest to de Sitter space. A small observed value of vacuum (or
dark) energy may be possibly explained by adjustment mechanism
realized by a scalar field (for a recent review see
ref.~\cite{Nobb}), or by quantum instability of de Sitter space
\cite{Two}.

On the other hand, considerable progress in quantum field theory
in curved space-time has been achieved in the well suited arena of
 de Sitter space-time, where calculations can be carried out
explicitly with different techniques \cite{Dowker}. In particular,
scalar fields were studied in detail because of their simple
properties.

Massless minimally coupled scalar field has the same description
of physical modes as gravitons in transverse-traceless gauge. This
relation allows to model infrared problems of quantum
gravity~\cite{Ford}, which may generate space-time
instability~\cite{Higuchi}, in terms of scalar fields. Appearance
of instability means that de Sitter invariance may be
spontaneously broken, and thus the true vacuum is not invariant
under the full symmetry group \cite{Mottola}; this in turn may
lead to the adoption of vacua which are invariant only under a
subgroup of the de Sitter group \cite{Kirsten}.

In this work we calculate vacuum expectation value of the operator
of scalar field squared, $\langle \varphi^2 \rangle$, in de Sitter
background, both in massive and massless case. Though this
quantity has been calculated in several papers there is still some
confusion about its infrared properties. Instead of relying on
rather {\it ad hoc} infrared cut-off we have derived an equation
governing evolution of $\langle \varphi^2 \rangle$ which is
solely based on de Sitter invariance. We have found the standard
expression for the massive case, while for massless field our
result for large time differs by sign from the previously
published ones.

This paper is organized as follows. In section II de Sitter
geometry is reviewed. In section III field quantization in
this space-time is discussed. In section IV a scalar field
$\varphi$ coupled to gravitation is considered. In section V an
equation describing the quantum average of $\varphi^2$ is derived
and its solutions are found both for massive and massless cases
with an emphasis on the minimally coupled field.  The conclusion
is presented in section VI.

Below the following notations are used. An overdot means
derivative with respect to time, the index "0" means time
component of tensor or vector, a letter from the middle of the
Latin alphabet, such as $i,j,k,\dots,$ means spatial components of
tensor or vector. The system of units is $c=\hbar=k=1$ and
$m_{Pl}^2/8\pi = 1$.

\section{De Sitter space-time}

The equations of gravitational field (the Einstein's equations)
with non-zero cosmological constant $\Lambda$, given the
gravitational action functional $S_{g}=-\int d^{4}x
\sqrt{-g}(R+2\Lambda)/2 $, have the form: \be \label{Ein}
R_{ab}-\frac{1}{2}\,R\,g_{ab}-\Lambda\,g_{ab}=T_{ab}, \ee where
$R$ is the scalar curvature, $g$ is the determinant of metric
$g_{ab}$ and $T_{ab}$ is the energy-momentum tensor of matter
fields.

 A very simple but non-trivial case is represented
by space-times with a constant curvature, locally characterized by
the condition $R_{abcd}=R(g_{ac}\,g_{bd}-g_{ad}\,g_{bc})/12 $,
which is equivalent to having a zero Weyl tensor $C_{abcd}$,
indicating that the space-time is conformally flat.

These space-times can be viewed as empty spaces with
$\Lambda=-R/4$ or as filled with a perfect fluid with the equation
of state $\varrho=-\,p$. The space with zero curvature $R=0$ is
Minkowski space-time: it is flat four-dimensional hyperplane. The
space with negative constant curvature is de Sitter space-time,
whose topology is $\mathcal{R}^{1}{\times} S^{3}$, while the one
with $R>0$ is anti-de Sitter space-time. The former can be viewed
as a four-dimensional hyperboloid embedded in five-dimensional
Riemannian space: in cartesian five-dimensional coordinates
$x^{a}$, with the metric
$\eta_{ab}=\textrm{diag}(+1,-1,-1,-1,-1)$, its points must satisfy
the relation: \be \label{5d}
x_{0}^{2}-x_{1}^{2}-x_{2}^{2}-x_{3}^{2}-x_{4}^{2}= -H^{-2}. \ee
Different coordinate systems may be used to describe de Sitter
space-time, this freedom corresponds to different choices of
Cauchy hyper-surfaces of constant time.

In flat coordinates the metric can be written in the
Friedmann--Robertson--Walker form: \be \label{dsmetric}
ds^{2}=dt^{2}-a^{2}(t)\,d \textbf{r}^{2}, \qquad\textrm{with}\quad
a(t)=\exp(Ht),\quad H=\sqrt{\Lambda/3}. \ee

The isometry group of de Sitter space-time is $SO(4,1)$, that is
the Lorentz group in a five-dimensional universe with four spatial
dimensions.

Actually $SO(4,1)$ is only one of the four disconnected parts of the
full symmetry group $O(4,1)$ \cite{Mottola}, the other parts are
obtained reflecting time, space, and both of them. The antipodal
transformation which is an element of the second part sends a point
$x$ to its antipodal point $\overline{x}$, lying in the opposite side
of the whole de Sitter space-time (for a discussion of antipodal
points, topology and symmetry see \cite{Norma,Innes}). If $X^{a}(x)$
is a five-vector locating $x$ in the metric $\eta_{ab}$, then
$\overline{x}$ is located by $-X^{a}(x)$.

Matter fields in general relativity are analyzed under some energy
conditions, among which we mention the assumed weak energy
condition, stating that at each space-time point $x$ there must be
$T_{ab}\,W^{a}\,W^{b}\geq0$ for any time-like vector $W
\,\in\,\mathcal{T}_{x}$ and by continuity for any null vector $W
\,\in\,\mathcal{T}_{x}$ too ($\mathcal{T}_{x}$ being the space of
tangent vectors at $x$) \cite{HE}.

In de Sitter space-time the Ricci tensor and the scalar curvature
have a very simple form: \be \label{Ricci}
R_{ab}=-3H^{2}\,g_{ab},\qquad R=-12H^{2}. \ee

From the Einstein equations follows the conservation law for the
energy-momentum tensor that can be written, in the chosen metric,
as: \be \label{r+p}
\nabla_{a}\,T^{a}_{\phantom{a}0}=\dot{\varrho}+3H(\varrho+p)=0.
\ee \emph{Thus in the exact de Sitter state, implying $\varrho +
\,p =0$, energy density $\varrho$ must be constant.}

\section{Quantum field theory in de Sitter space.}

The choice among different possible quantization procedures must
both take into account their consequences for symmetry breaking,
 anomalies and zero point energy, and the
representations of the symmetry group in question. Scalar
representations of de Sitter group can be divided into the
principal series $(m^{2}\geq 4H^{2})$, the complementary series
$(0<m^{2}<4H^{2})$, and the discrete series $(m^{2}=0$ is the only
interesting case) \cite{Tolley}.

For massive scalar fields, corresponding to the first two cases,
several different procedures are known which lead to a well
defined Fock space. Massless case is more cumbersome, because, as
it has been shown by several authors \cite{Mottola}, there is no
Fock space corresponding to a de Sitter invariant vacuum. This
consideration leads to the Gupta--Bleuler quantization (naturally
extensible to the massive case) in order to avoid such problem and
keep intact the full de Sitter invariance \cite{Bievre}.

A rigorous covariant quantization is achieved in the global
approach to quantum field theory \cite{Dewitt}, where the starting
point is the local symmetry and de Sitter group is considered as a
charge group generated by the Killing vector fields, arising in
the treatment of the gravitational field as an external field in
the frameworks of the background field method.

An axiomatic approach, generalizing to curved space the axiomatic
methods of Minkowskian quantum field theory \cite{Streater},
impose some basic requirements on the two-point functions which
are necessary to obtain a well defined quantum theory
\cite{Moschella}. Reasonable general principles seem to be
covariance, locality, and positive definiteness of the two-point
functions, as hold in Minkowskian case. The spectral condition,
instead, cannot be literally translated \cite{Epstein}, and one
must adopt a weaker one, based on the property that the two-point
function has to be a boundary value of an analytic function with
the correct $i\epsilon$ prescription. The latter is also a
generalization of the Hadamard condition \cite{Kay}. Being
formulated only for free field theory, this condition may be
considered as a consequence of the Einstein's equivalence
principle. It assures that Green functions have the same
singularities as in flat space.

Causal problems arise from antipodal points, where Green functions are
also singular, although
singularities occurring at those points are unobservable, since such
points are always separated by a horizon \cite{Bousso}. Instead of implying
global condition on the propagators, one can avoid this problem in the
flat coordinate system, where a half of the space, containing antipodal
points, is excluded.

Below the standard quantization procedure will be assumed, with
the usual commutation relation between the scalar field and its
conjugate momentum, which can be recast into the usual commutation
relations among creation and annihilation operators, provided the
Wronskian condition on the mode functions holds~\cite{BD}. There
is a close resemblance to the flat space-time case, the only
difference being in the explicit form of the mode functions. In
de Sitter space the latter are not plane waves, but a linear
combination of the Hankel functions, with two constants
coefficients, determined by the Wronskian condition plus the
choice of the vacuum state.

Actually there is a one-parameter family of de Sitter invariant
vacua~\cite{Mottola}, called \emph{$\alpha$-vacua}. The best known
and often used one is determined by the Bunch--Davies
prescription, leading to the usual Euclidean vacuum in the
limiting case of flat space-time. Since the discrete PT (parity
and time reversal) symmetry is a subgroup of de Sitter group,
the condition of invariance with respect to the total group
automatically makes the free vacuum propagator PT invariant.
Supplementing this choice with the above mentioned Hadamard
condition leads to the \emph{Euclidean} or Bunch--Davies vacuum
\cite{Mottola,BD}.

It can be shown~\cite{Boer} that among $\alpha$-vacua in planar
coordinates the Euclidean vacuum is naturally interpreted as the
state with no particles on the horizon at $t=\infty$
\cite{Spradlin}. Another feature of the Euclidean vacuum is that
$\alpha$-vacua in an interacting theory are ill-defined
\cite{Einhorn}, but can be regarded at least approximately as
excited states in the Euclidean vacuum \cite{Larsen}.
Renormalization considerations also lead to the conclusion that
the Euclidean vacuum is the only physically acceptable state
\cite{Banks}.

One more indication of the physical relevance of the Euclidean
vacuum comes from the study of the behavior of the energy-momentum
tensor. It has been proved \cite{Habib} that the renormalized
value of this tensor for all ultraviolet and infrared physically
allowed initial states asymptotically approaches the one obtained
in the Bunch--Davies state. The renormalized value has been
calculated by the adiabatic regularization. Although this scheme
is not covariant, in the Friedmann--Robertson--Walker space-times
it is equivalent to the covariant point-splitting procedure.

Through back-reaction of the expectation value of the
energy-momentum tensor on gravity its quantum fluctuations could
influence the metric. Since for massless free fields $T_{ab}$
contains only derivatives of the field, it is not sensitive to
long-wavelength modes. That is why in non-interacting theory it
does not manifest breaking of the full de Sitter group. Purely
gravitational back-reaction mechanism, instead, could drive
changes in the metric, terminating the de Sitter phase: this might
be a viable way to exit inflationary era (see \cite{wood,Brandi}
and references therein).

It was proved \cite{Mottola} that in massless case there is no Fock
space for a full de Sitter invariant propagator, because the calculation
in the limit $m\rightarrow 0$ are performed through an infrared cutoff,
which excludes zero modes from the integration over the phase space and
hence the set of mode functions is not complete. The nature of
those cutoffs is not clear. It may be understood through reasonable
physical arguments, for example from the size of the initial
wedge that inflated \cite{Linde}.
Infrared properties of massless self-interacting scalar field have been considered in \cite{Onemli}.

Particle interpretation requires a  Fock space, that is why some
authors \cite{Kirsten},\cite{AF} postulated less invariant vacua,
for example compact $O(4)$ invariant vacua or non-compact $E(3)$
invariant vacua. In the former case the compactness allows to
construct new zero modes in order to have a complete set of mode
functions. In the $E(3)$ invariant vacua this is not possible. For
this reason such vacua are considered as idealizations of physical
vacua.

Instability of de Sitter space-time in the presence of massless
minimally coupled scalar field forces one to rely on the
Allen--Folacci vacuum \cite{AF}, which is not invariant under the
full symmetry group. This may be seen, as in the massive case,
through the energy-momentum tensor: its renormalized value in the
Allen--Folacci vacuum is an attractor solution for any other
vacuum choice in the massless case \cite{Habib}. In the
non-invariant vacua the expectation value of the energy-momentum
tensor becomes different from the one obtained in the Euclidean vacuum. It can
also depend on time \cite{Kirsten}, leading to possible changes in
the metric through the back-reaction mechanism.

\section{Scalar field}

A scalar field $\varphi$ with mass $m$ and coupling $\xi$ to the
curvature scalar, is described by the action: \be \label{ac}
S=\frac{1}{2} \int d^{4}x \sqrt{-g} \left[\nabla_{a}\,
\varphi\nabla^{a}\,\varphi - m^{2}\varphi^{2} - \xi
R\,\varphi^{2}\right]. \ee which leads to the following equation
of motion \be \label{em} \nabla_{a}\,\nabla^{a}\,\varphi +
m^{2}\varphi + \xi R\,\varphi
 = 0.
\ee

In metric (\ref{dsmetric}) the equation of motion reads:
\be
\label{emds} \ddot\varphi - \left(\triangle/a^2\right)\,\varphi +
3H\dot\varphi + m^{2}\varphi + \xi R\,\varphi = 0, \ee
where
$\triangle$ means the tridimensional Laplacian operator in flat
space.

The energy-momentum tensor of the scalar field $\varphi$ is
defined as \be \label{emt} T^{ab}\equiv
\frac{-2}{\sqrt{-g}}\frac{\delta S}{\delta g_{ab}}. \ee

Remembering that $g_{ab}\,g^{bc}=\delta_{a}^{\phantom{a}c}$ and the matrix
property ${\rm tr}\,\ln\,g_{ab}=\ln\,\det\,g_{ab}$ and taking the
infinitesimal
variation of the former one can obtain the following useful identities:
\begin{equation*}
\label{var}
\frac{\delta\,\sqrt{-g}}{\delta\,g_{ab}} =
\frac{1}{2}\sqrt{-g}\,g^{ab},\quad\quad
\frac{\delta\,g^{cd}}{\delta\,g_{ab}} =-g^{ac}\,g^{bd},\quad\quad
g^{cd}\,\frac{\delta\,R_{cd}}{\delta\,g_{ab}}  =
\nabla^{a}\nabla^{b}-g^{ab}\,\nabla^{e}\,\nabla_{e}.
\end{equation*}

At this point the expression for the operator of the
energy-momentum tensor can be easily derived:
\begin{align}
\label{T} T^{ab} = &-\frac{1}{2}\,
g^{ab}\,\left(\nabla_{c}\,\varphi\, \nabla^{c}\,\varphi -
m^{2}\varphi^{2}\right) +\nabla^{a}\,\varphi\,
\nabla^{b}\,\varphi -\xi\left(R^{ab}- \frac{1}{2}\,g^{ab}\,
R\right)\,\varphi^{2} \nonumber\\
 & -\xi\left(\nabla^{a}\,\nabla^{b}\,-g^{ab}\,\nabla_{c}\,
\nabla^{c}\right)\varphi^{2}.
\end{align}
As expected this tensor is covariantly conserved: \be \nabla_a
T^{a}_{\phantom{a}b} = 0 .\label{conserv-T} \ee Let us note that
the conservation is realized only on solutions of the equation of
motion (\ref{em}).

It's straightforward to write the operator expression for the
trace of $T_{ab}$: \be \label{trace} T^{a}_{\phantom{a}a}=
-\nabla_{a}\,\varphi\,\nabla^{a}\,\varphi+2\,m^{2}\,\varphi^{2}+\xi\,
R\,\varphi^{2}+3\,\xi\,g^{ab}\,\nabla_{a}\,\nabla_{b}\,
\varphi^{2}. \ee In the special case of a massless conformally
coupled scalar field $(m=0$, $\xi=1/6)$ this expression vanishes
on solutions of the equation of motion \eqref{em}. However, since
the product of two operators in coinciding space-time points is
ill-defined a regularization prescription is necessary. The latter
may break the classical identities and lead to well known quantum
anomalies. In particular, the vacuum expectation value of the
trace of $T_{ab}$ possesses such an anomaly called the \emph{trace
anomaly} or \emph{conformal anomaly}. Independently on
renormalization technique the resulting expression for the anomaly
in four dimensions is \cite{BD,Dewitt}: \be \label{anomaly}
\langle T^{a}_{\phantom{a}a}\rangle_{ren}=
-\frac{1}{2880\,\pi^{2}}\left(R_{abcd}\,R^{abcd}-R_{ab}\,R^{ab}-\Box\,R\right).
\ee In de Sitter space it becomes: \be \label{dsanomaly} \langle
T^{a}_{\phantom{a}a}\rangle_{ren}= \frac{H^4}{240 \pi^2} . \ee

The complete expression for the regularized vacuum expectation
value of $T_{ab}$ in de Sitter space for arbitrary $m$ and $\xi$
was calculated in ref.~\cite{Dowkercri,Bunch} and reads:
\begin{align}
\langle T_{ab}\rangle_{ren} & = \frac {g_{ab}}{64\pi^2} \left\{
m^2\left[m^2 - \left( \xi - \frac{1}{6} \right)R\right] \left[\psi
\left(\frac{3}{2} + \nu \right) + \psi \left(\frac{3}{2} - \nu
\right) - \ln \left(\frac{12m^2}{|R|} \right)\right]\right.
\nonumber \\
& \left. -\frac{1}{2}\left(\xi - \frac{1}{6} \right)^2 R^2
+\frac{R^2}{2160} - m^2 \left(\xi - \frac{1}{6} \right) R -
\frac{m^2 R}{18} \right\}, \label{T-ab-tot}
\end{align}
where $\psi$ is the logarithmic derivative of the Gamma-function
and \be \nu^2 = \frac{9}{4} + 12 \left( \frac{m^2}{R} - \xi
\right) .\label{nu} \ee

The result (\ref{T-ab-tot})
is obtained with a covariant regularization procedure which
respects conservation condition (\ref{conserv-T}) and de Sitter
symmetry according to which $T_{ab} \sim g_{ab}$.

In the conformal case, $m=0$ and $\xi =1/6$, the trace of
(\ref{T-ab-tot}) coincides, as expected, with (\ref{dsanomaly}).
In the limit of $m\rightarrow 0$ and minimal coupling to gravity,
$ \xi = 0$, the energy-momentum tensor becomes: \be T_{ab} =
\frac{g_{ab} H^4}{\pi^2} \left( \frac{3}{32} + \frac{1}{960} -
\frac{1}{32} \right) . \label{T-m0-xi0} \ee The first term in the
brackets comes from $m^2 R\, \psi(3/2 -\nu)/6$ in eq.
(\ref{T-ab-tot}), simply using the definition
$\psi(z)=d(\ln\Gamma(z))/dz$ and the identity
$\Gamma(z+1)=z\Gamma(z)$ in the limit $z\rightarrow 0$. It is the
standard non-anomalous term in the energy density at $m=0$. The
other two terms come from anomalous contributions: the first of
them is purely anomalous one which survives in conformal limit,
while the second disappears only for $\xi =1/6$ and is
non-vanishing for $m=0$ and $\xi=0$.

\section{Calculation of quantum average of $\varphi^2$ in de Sitter space.
}

The quantity $\langle\varphi^{2}\rangle$, that is the quantum
average value of the product of the scalar field operators
$\hat{\varphi}$ in coincident space-time points, plays a primary
role in quantum field theory. There are different methods to
obtain its expression, which is divergent without an appropriate
renormalization of the field operator  $\hat{\varphi}$.

Such methods \cite{Ball} comprise the use of zeta function
techniques, point-splitting regularization, dimensional
regularization, etc. giving rise in general to different
renormalized value of $\langle\varphi^{2}\rangle$. The usual
expansion of the field  operator $\hat{\varphi}$ through its mode
functions, that is through the Hankel functions in de Sitter
space, allows to find an expression for
$\langle\varphi^{2}\rangle$ in terms of Digamma and Hypergeometric
functions, which gives the result \cite{Vford} (originally
obtained with point-splitting regularization): \be \label{bd}
\langle\varphi^{2}\rangle=\frac{1}{16\pi^{2}}
\left\{m^{2}\ln\left(\frac{\mu^2}{12m^{2}}\right)
+\left[m^{2}-\left(\xi-\frac{1}{6}\right)R\right]
\left[\ln\left(-\frac{R}{\mu^{2}}\right)+\psi\left(\frac{3}{2}+\nu\right)
+\psi\left(\frac{3}{2}-\nu\right)\right]\right\} . \ee The
renormalization mass $\mu$ is to be chosen in such a way that
$\langle\varphi^{2}\rangle $ vanishes in flat space limit i.e.
when $R\rightarrow 0$. In massive case it can be achieved with
$\mu^{2}=12 m^{2}$. Indeed, when $m/H$ is large then $\nu \approx
im/H$ and \be \psi(3/2 +\nu) + \psi (3/2 -\nu) \approx 2\ln (m/H).
\label{re-psi} \ee This term cancels down with $\ln (-R/\mu^2)$ if
$\mu^2 = 12 m^2$. In massless case and vanishing
 $R$ $\langle\varphi^{2}\rangle \rightarrow 0$ for
arbitrary value of $\mu^{2}$ and almost any $\xi$. However, for
$\xi =0$ the limit $m\rightarrow 0$ is singular and \be
\langle\varphi^{2}\rangle \rightarrow \frac{3H^4}{8\pi^2 m^2} .
\label{phi2-m0} \ee In this case the transition to $H=0$ is
non-trivial, if first $m \rightarrow 0$.

In this work we will develop a different approach to calculation
of the quantum average value of $\varphi^2$ in de Sitter space. We
will derive an ordinary differential equation governing the
evolution of $f(t) \equiv \langle\varphi^{2}\rangle $. To this end
we will use the equation of motion for the quantum operator
$\hat{\varphi}$ (\ref{emds}) and the commutation of the quantum
averaging and differentiation. This approach is similar to that
indicated in ref.~\cite{Vilenkin}. The scalar field
$\hat{\varphi}$ is an operator-valued distribution, and the rule
for derivatives of distributions allows to use the standard rule
for derivatives of the product of functions; in our case it is
$\hat{\varphi}{\cdot} \hat{\varphi}$. In this way we obtain: \be
\label{d2} g^{ab}\,\nabla_{a}\,\nabla_{b}\,f=
g^{ab}\,\nabla_{a}\langle 2\varphi\,\nabla_{b}\,\varphi \rangle =
\langle 2\,g^{ab}\,\nabla_{a}\,\varphi\,\nabla_{b}\,\varphi+
2\,\varphi\, g^{ab}\,\nabla_{a}\,\nabla_{b}\,\varphi\rangle . \ee

This identity permits to write the following equation for $f$: \be
\label{emf} (\nabla_{a}\,\nabla^{a}+m^{2}+\xi\,R)f= 2\,\langle
\nabla_{a}\,\varphi\,\nabla^{a}\,\varphi \rangle - m^{2}f-
\xi\,R\,f. \ee To make this equation meaningful we need to
calculate the first term in the r.h.s. of (\ref{emf}) in terms of
known quantities. Expressing it through the trace of the
energy-momentum tensor \eqref{trace}, and assuming unbroken de
Sitter invariance, one can get a solvable closed equation for $f$.
After straightforward algebra one obtains: \be \label{eq}
(1-6\,\xi) \nabla_{a}\,\nabla^{a}\,f-2\,m^{2}f = -2\langle
T^{a}_{\phantom{a}a}\rangle = -8\,\langle\varrho\rangle, \ee where
$\langle \varrho \rangle$ is the vacuum energy density of quantum
fluctuations of $\varphi$. If de Sitter invariance is unbroken
then $\langle\varrho\rangle = \langle
T^{a}_{\phantom{a}a}\rangle_{ren}/4$ and the trace can be
trivially calculated from eq.~(\ref{T-ab-tot}) for arbitrary
values of $m$ and $\xi$.

In homogeneous background $f$ should depend only on time and the
equation can be rewritten as: \be \label{eq1}
(1-6\xi)\left(\ddot{f}
+3\,H\,\dot{f}\right)-2\,m^{2}f=-8\,\langle\varrho\rangle . \ee
Because of de Sitter invariance $\langle\varrho\rangle =$
constant, and this equation can be solved explicitly. In the case
of minimal coupling to gravity, i.e. $\xi =0$, the solution is:
\be \label{mxi0-1}
f=C_{1}\exp{\left[\left(-\frac{3}{2}\,H-\sqrt{\frac{9}{4}\,H^{2}+
2m^{2}}\right)t\right]}+C_{2}\exp{\left[\left(-\frac{3}{2}\,H+
\sqrt{\frac{9}{4}\,H^{2}+2m^{2}}\right)t\right]}+
\frac{4\langle\varrho\rangle}{m^{2}}\;, \ee where $C_1$ and $C_2$
are some numerical constants. Surprisingly the solution is time
dependent if any of $C_{1,2}$ is non-zero, moreover it contains an
exponentially rising term, proportional to $C_2$. The solution is
constant, $f = 4 \langle \rho \rangle /m^2$, independently of
initial conditions only for conformal coupling, $\xi = 1/6$. It is
unclear if time dependence or, at least, exponential rise can be
killed by an appropriate choice of physically justified initial
conditions. The limit of zero mass indicates the opposite, that
$C_2$ should be non-zero (see the end of this Section). Possibly
fast (exponential) evolution of $\langle\varphi^{2}\rangle$ can be
helpful for solution of the well known problem of vacuum energy.

In the case ${9}H^{2}/4\gg 2m^{2}$ and $t< 3H/2m^2 $ the solution
is \be \label{xi0} f  =
C_{1}\exp{(-3\,H\,t)}+C_{2}\left(1+\frac{2m^2}{3H}\,t \right)
+\frac{4\langle\varrho\rangle}{m^{2}}. \ee If $C_1$ and $C_2$ are
not singular at $m=0$, the dominant term for $m\rightarrow 0$ is
\be \langle \varphi^2 \rangle =
\frac{4\langle\varrho\rangle}{m^{2}}. \label{m-tend-0} \ee It is
formally the same as that found for $\xi = 1/6$ above but one
should keep in mind that $\langle \rho \rangle$ depends both on
$m$ and $\xi$. Expression (\ref{m-tend-0}) would agree with the
previously established one~\cite{Vford,Starobinski} if we
substitute for the energy density of quantum fluctuations of
$\varphi$ the small-mass limit of the 
result~\cite{Dowkercri,Bunch}:
$\langle\varrho\rangle={3H^{4}}/({32\pi^{2}})$, i.e. only the
first term in eq.~(\ref{T-m0-xi0}), while the anomalous
contributions are disregarded. An account of the anomaly changes
the numerical coefficient and $\langle \varphi^2 \rangle =
3H^4/8\pi^2m^2$ turns into $\langle \varphi^2 \rangle =
61H^4/240\pi^2m^2$.

However, it may be not as simple as that because, taken as it is,
eq. \eqref{eq1} is not consistent in conformal limit, $m=0,
\xi=1/6$. Indeed, in this limit the l.h.s. of this equation
vanishes, while the r.h.s. is non-zero due to trace anomaly. A
possible way out is a singularity in
$\langle \varphi^2 \rangle$ at $m=0$ or $\xi=1/6$, as indicated by eqs. \eqref{mxi0-1},\eqref{m-tend-0} according to which $\langle \varphi^2 \rangle\sim \langle \varrho \rangle/m^2$.

One can also solve eq. (\ref{eq}) directly for $m=0$ (and $\xi
=0$): \be \label{mxi0} f=C_{1} + C_{2}\,e^{-3H\,t}-
\frac{8}{3}\,\frac{\langle\varrho\rangle}{H}\,t, \ee which gives
the late time behavior \be \label{ltm0} f\approx
-\frac{8}{3}\,\frac{\langle\varrho\rangle}{H}\,t. \ee This
solution almost coincides with the earlier found
one~\cite{Vford,Starobinski} - the absolute value is the same if
anomaly is not included but, surprisingly, the sign is opposite.
The same solution \eqref{mxi0} can be obtained from expression
(\ref{xi0}) in the limit of zero mass under condition that the
solution is not singular at $m=0$. To realize that the coefficient
$C_2$ should be non-vanishing and singular in $m$: $C_2 = C_{20} -
4\langle\varrho\rangle/m^2$, where $C_{20}$ is a non-singular
constant at $m=0$. With $m$ tending to zero the last term cancels
out the singular in $m$ part in eq. (\ref{xi0}) and the remaining
one coincides with (\ref{ltm0}).

We see that for small mass and large time the average value of
$\varphi^2$ becomes negative. It looks strange because $\varphi^2$
is a positive definite operator. Still it may be true because the
vacuum expectation value of this operator needs to be renormalized
and after (infinite) renormalization 
can become negative.

\section{Conclusion}

The case of spontaneous symmetry breaking in de Sitter space-time
is similar to that of two-dimensional flat space-time with a
broken Lorentz invariance \cite{FV86}. Although it has been
proposed in ref.~\cite{Ratra}, on the basis of functional methods,
that in de Sitter space-time symmetries are always dynamically
restored, this is probably not the case.

The occurrence of this phenomenon may be connected with the choice
of the vacuum state \cite{Polarski} and related to this choice
character of the infrared behavior in the massless limit. It also
relies on the assumptions imposed on admissible physical states
and theoretical ambiguities in quantum field theory in curved
space. The properties of the admitted states, as e.g. behavior at
large distances, zero modes, or possible interactions, determine
the character of infrared divergences or directly the symmetry
breaking \cite{Redmount}. In any case infrared divergences could
quite naturally lead to an instability. Possible manifestation
of this phenomenon may be realized by
quantum gravity effects in de Sitter space. They may lead to quantum
instability of this state \cite{Woody,Witten}. In other words, due to
quantum effects de Sitter solution could tend to some other solutions
of the Einstein equations loosing de Sitter symmetry~\cite{Folacci}.

In this paper we have derived and solved a differential equation
which describes the behavior of the vacuum expectation value of
the quantum operator product $\varphi^2$ in de Sitter space. We
have found that generally the solution is time dependent, though
for non-zero mass a very special constant solution exits. However
this solution cannot be continuously transformed to massless
limit.

Our result is not based on any \emph{ad hoc} infrared cutoff and
it agrees with the standard one in the case of small but
non-vanishing mass. This agreement is an indication for the
validity of our approach. In the massless case, however, we obtain
a surprising difference - the magnitude of $\langle \varphi^2
\rangle$ is the same as found in the earlier papers but the sign
is opposite. The large time behavior of this solution does not
depend upon initial conditions.

We would like to stress that the result is obtained under
assumption of de Sitter symmetry for quantum expectation values,
which demands $\varrho =$ constant and $\langle\varrho\rangle +
\langle p\rangle =0$. The numerical coefficient in eq.
\eqref{ltm0} is determined by the expression \cite{Dowkercri,Bunch} for
$\langle\varrho\rangle$. Possibly time dependence of quantum
fluctuations of scalar field is an indication of spontaneous
symmetry breaking.

The minus sign in the massless case with respect to other
calculations could be a signal of a quantum infrared anomaly in de
Sitter space. On the other hand, this discrepancy could emerge as
an indication of some problem in the limiting procedure to get
zero mass from the massive case. This situation has a close
resemblance to the appearance of the conformal anomaly in the
massless limit of the renormalized value of the trace of the
energy-momentum tensor \eqref{trace}. After the preparation of
this work we became aware of a recent paper about gravitons in de
Sitter space \cite{Marozzi}, where the renormalized value of the
energy-momentum tensor has the sign opposite to that in the
literature.

Further investigations may clarify the situation, in particular a
step by step comparison with results which can be obtained in the
Euclidean sector which, although not shown here, would present the
same anomalous behavior in the massless case. An investigation of
Euclidean case may be helpful in understanding where a possible
breakdown of the standard behavior can occur.

\acknowledgments

A.D. Dolgov is grateful to M. Einhorn, E. Mottola and R. Woodard
for discussion at an earlier stage of this work. The work of D.N.
Pelliccia has been supported by INFN grant n.10271. He is grateful
to F. Zaccaria and G. Esposito for encouragement, to M. Giannotti
and I. Parenti for useful discussions. We thank A. Vilenkin for
comments.

\end{document}